\documentclass[preprint,12pt]{elsarticle}
\usepackage{amsmath,amsfonts}
\usepackage{algorithmic}
\usepackage{algorithm}
\usepackage{array}
\usepackage[caption=false,font=normalsize,labelfont=sf,textfont=sf]{subfig}
\usepackage{textcomp}
\usepackage{stfloats}
\usepackage{url}
\usepackage{soul} 
\soulregister\cite7
\soulregister\ref7 
\soulregister\bf7 
\usepackage{color, xcolor}
\usepackage{hyperref} 
\hypersetup{hypertex=true,
            colorlinks=true,
            linkcolor=black,
            hidelinks
            }
\usepackage{verbatim}
\usepackage{graphicx}
\usepackage{color}

\journal{Knowledge-Based Systems}

\begin{document}

\begin{frontmatter}

\title{Cross-modal Generative Model for Visual-Guided Binaural Stereo Generation}

\author{Zhaojian~Li, Bin~Zhao and Yuan~Yuan\corref{mycorrespondingauthor}}
\cortext[mycorrespondingauthor]{Corresponding author}
\address{School of Computer Science and Artificial Intelligence, OPtics and ElectroNics (iOPEN), Northwestern Polytechnical University, Xi’an 710072, Shaanxi, PR China}

\begin{abstract}
Binaural stereo audio is recorded by imitating the way the human ear receives sound, which provides people with an immersive listening experience.
Existing approaches leverage autoencoders and directly exploit visual spatial information to synthesize binaural stereo, resulting in a limited representation of visual guidance.
For the first time, we propose a visually guided generative adversarial approach for generating binaural stereo audio from mono audio.
Specifically, we develop a Stereo Audio Generation Model (SAGM), which utilizes shared spatio-temporal visual information to guide the generator and the discriminator to work separately.
The shared visual information is updated alternately in the generative adversarial stage, allowing the generator and discriminator to deliver their respective guided knowledge while visually sharing.
The proposed method learns bidirectional complementary visual information, which facilitates the expression of visual guidance in generation.
In addition, spatial perception is a crucial attribute of binaural stereo audio, and thus the evaluation of stereo spatial perception is essential.
However, previous metrics failed to measure the spatial perception of audio.
To this end, a metric to measure the spatial perception of audio is proposed for the first time.
The proposed metric is capable of measuring the magnitude and direction of spatial perception in the temporal dimension.
Further, considering its function, it is feasible to utilize it instead of demanding user studies to some extent.
The proposed method achieves state-of-the-art performance on 2 datasets and 5 evaluation metrics. 
Qualitative experiments and user studies demonstrate that the method generates space-realistic stereo audio.
\end{abstract}


\end{frontmatter}

\section{Introduction}
\begin{figure}[t]
  \centering
   \includegraphics[width=1\linewidth]{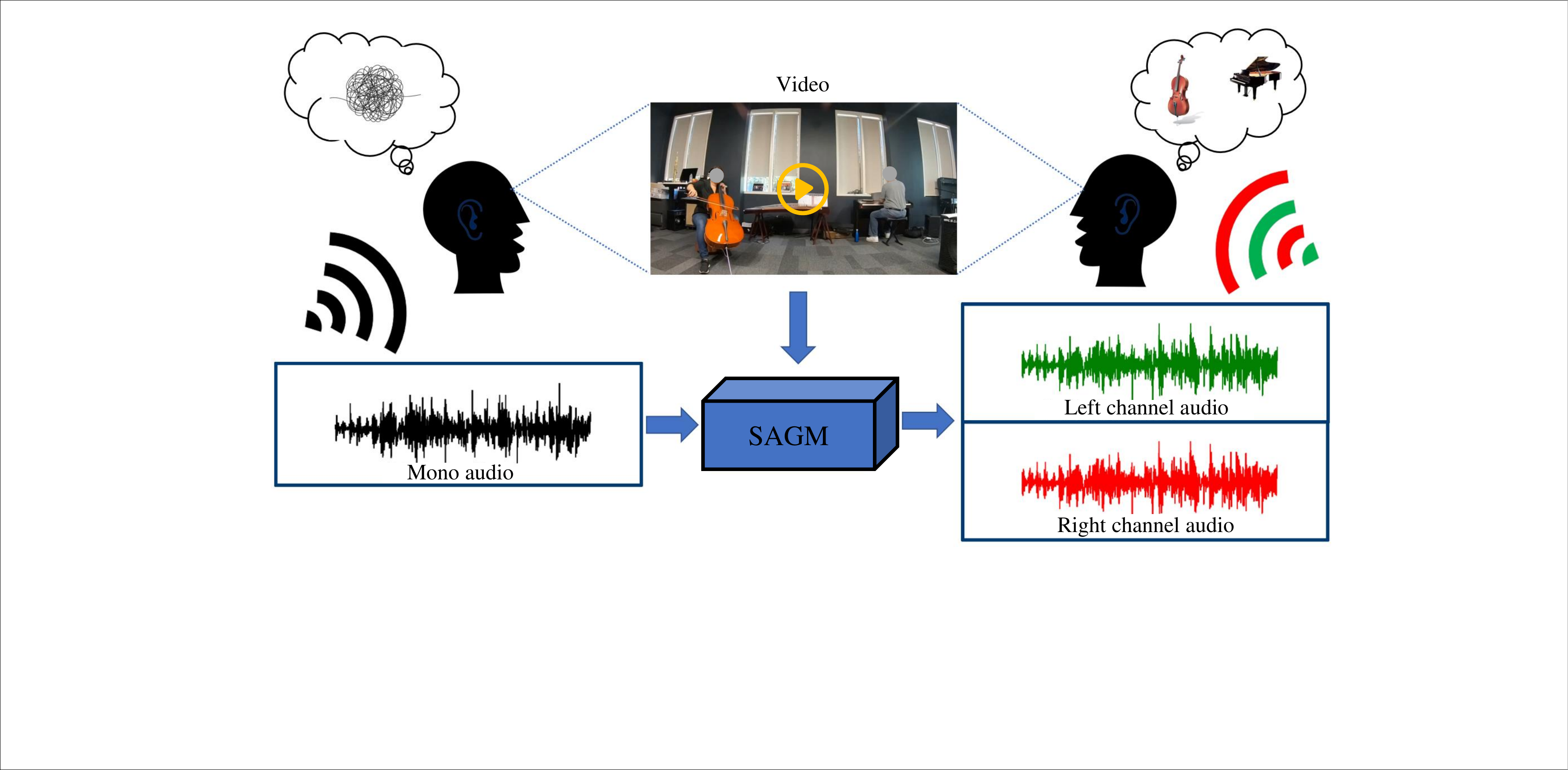}

   \caption{SAGM generates binaural stereo audio from mono audio guided by visual information. 
   Mono audio lacks spatial location information of the sounding object. 
   The proposed method can reproduce the spatial perception of sound, transporting the listener back to the sound scene at the time of recording.
   }
   \label{fig:onecol}
\end{figure}

The binaural structure of the human head makes the sound heard have a binaural effect, allowing people to estimate the direction of a sound \cite{wu2018audio, DBLP:journals/taslp/PanZWWL21, zhou2020survey} by relying on the volume difference, time difference, and timbre difference between the two ears \cite{chen2021multimodal}.
Interaural Time Differences (ITDs) and Interaural Level Differences (ILDs) are the key factors in the formation of the binaural effect.
ITDs is the time difference between the same sound reaching both ears, while ILDs is the amplitude difference between them \cite{DBLP:conf/iccv/RachavarapuAS021}.
Much behavioral and psychoacoustic evidence \cite{rosen2022effect, li2022precedence} has confirmed that ITDs and ILDs play an important role in inferring the location of sound sources \cite{ning2022audio, zhou2023exploiting, mohamed2022face}. 
However, the lack of spatial knowledge of the sound source in mono audio \cite{middya2022deep} results in our inability to distinguish the direction of the sound source (see Fig. \ref{fig:onecol}).

Compared with the directionality of visual sensation, binaural auditory perception provides people with a surround understanding of space \cite{DBLP:conf/iclr/RichardMGKBTS21}.
For this reason, the generation of realistic stereo sound is essential for people to immerse themselves in artificial spaces. 
Binaural stereo audio is recorded with an artificial or dummy heads plugged in with microphones, which simulates the way human ears receive sound to reflect human hearing experience in the real world \cite{hammershoi2002methods}. 
It is widely applicable in virtual reality (VR) \cite{li2020spatio, robotham2018evaluation}, augmented reality (AR) \cite{del2022hybrid, DBLP:journals/taslp/Ben-HurAMR21}, and games \cite{franvcek2023performance, patro2022explanation}.
A common scenario where stereo audio is employed to improve the auditory experience is when a movie theater arranges two speakers at the front left and right of the audience to reproduce the sound of the two channels.
As a result, the audience is able to gain a spatial perception of the sound while watching the animation.
In the recently booming metaverse field, stereo audio also has great application potential, which can provide people with auditory mapping of the real world in the virtual world.

Binaural stereo video brings an impressive audiovisual feast to people.
Nevertheless, there are much fewer binaural stereo videos than mono videos \cite{zhao2021RSGN, zhao2018hsa} because recording binaural stereo audio requires professional equipment and knowledge \cite{DBLP:conf/apccas/GSAAPS20}.
Moreover, the price of the equipment is relatively high \cite{DBLP:conf/cvpr/XuZ0D0L21}.
Simultaneously, numerous studies have demonstrated that the spatial information contained in stereo audio can improve the model's performance on certain tasks, such as sound source localization \cite{kim2023gaussian, DBLP:conf/cvpr/YangRS20}, utterance clustering \cite{DBLP:journals/corr/abs-2009-05076}, and speech separation \cite{DBLP:conf/ijcnn/GogateD0H20}.

Leveraging visual modalities to guide the generation of binaural stereo audio is a promising approach.
Therefore, visual modality, as the only source of knowledge for spatial perception, plays a crucial role in binaural audio generation.
Most of the existing approaches \cite{DBLP:conf/nips/MorgadoVLW18, DBLP:conf/cvpr/GaoG19, DBLP:conf/eccv/ZhouXLWL20} are all based on autoencoders and only consider the spatial information of the visual modality.
However, due to the construction of autoencoders, the direct utilization of visual information leads to the limited ability of visually guided learning.
In addition, we argue that temporal information is also crucial for the generation of stereo audio when the sounding object is in motion.
To this end, we propose a Stereo Audio Generation Model (SAGM) guided by a shared spatio-temporal visual modality to generate stereo audio.
Different from previous methods, the proposed method adopts a generative adversarial approach to introduce spatio-temporal visual information into the generator and discriminator simultaneously to provide targeted guidance for their generation and discrimination work, respectively.
For the generator, visual information is used to provide guided knowledge to help the generator reproduce the spatial perception of sound.
For the discriminator, the visual information provides a reference for the discriminator's decision. 
The visual information is alternately updated between the generator and the discriminator to transfer their respective learned guidance information. 
SAGM learns bidirectionally complementary visual guidance through generative adversarial interactions and generates more realistic stereo audio guided by shared visual information.

Our contributions in this paper are in three folds:

1) We propose a stereo audio generation model guided by a shared spatio-temporal visual modality to facilitate the expression of visual guidance.
The visual guidance is alternately updated during generative adversarial process to learn bidirectionally complementary visual knowledge.

2) We propose a novel metric that measures the magnitude and direction of spatial perception in the temporal dimension.
To our best knowledge, it is the first metric capable of evaluating spatial perception. 
It is feasible to replace laborious manual evaluation to a certain extent.

3) Extensive experimental results on two benchmark datasets demonstrate that the proposed method is superior to state-of-the-art methods and generates realistic binaural stereo audio.
Ablation experiments verify the significance of spatio-temporal visual guidance and visual sharing.

\section{Related Work}
\subsection{Audiovisual Source Separation}
The sight and sound of a sound source are often related because they share the same semantic, temporal, and spatial properties \cite{zhou2023seco, DBLP:conf/mm/ShuoJ0021, DBLP:conf/iccv/MajumderAG21, DBLP:conf/cvpr/GanHZT020, liVideoDistillation2021}.
Audiovisual source separation utilizes visual modality to separate individual sound source from a mixed sound source.
Afouras \emph{et al.} \cite{DBLP:conf/eccv/AfourasOCZ20} utilized a self-supervised learning method to train a model that can associate audio and video speaking objects. 
The whole model follows a two-stream structure.  
The finally extracted embedding can be used for some downstream tasks \cite{DBLP:conf/nips/FisherDFV00, DBLP:conf/interspeech/ChungNZ18, DBLP:conf/iccvw/RothXPCKMGKRSS19}, such as separation of sound from multiple speakers \cite{DBLP:journals/tog/EphratMLDWHFR18}. 
Unlike \cite{DBLP:conf/eccv/AfourasOCZ20} which requires an additional network, Zhao \emph{et al.} \cite{DBLP:conf/eccv/ZhaoGRVMT18} introduced PixelPlayer, an end-to-end system that utilizes a significant number of unlabeled videos to learn to locate the image area where the sound is generated and decomposes the input sound into a series of components that represent each pixel's sound. 
The experimental results on the MUSIC dataset \cite{DBLP:conf/eccv/ZhaoGRVMT18} demonstrate that the hybrid and separation framework proposed by the author is superior to multiple baselines in the task of sound source separation.
Owens \emph{et al.} \cite{DBLP:conf/eccv/OwensE18} proposed a pretext task \cite{DBLP:conf/iccv/ArandjelovicZ17, DBLP:conf/cvpr/YangRS20} that jointly models multi-perceptual representations of vision and sound to identify whether input video frames and corresponding audio are temporally synchronized.
The embedding learned by the model is applied to downstream tasks to realize sound source localization \cite{DBLP:journals/pami/SenocakOKYK21}, audiovisual activity recognition \cite{DBLP:conf/iccv/SubedarKLTH19}, and audio separation inside and outside the screen.
The motion cues of objects \cite{li2017multiview} can also guide sound source separation. 
Zhao \emph{et al.} \cite{DBLP:conf/iccv/ZhaoGM019} tried to explain the sound source from the visual movement of the object, and proposed a system that can capture the movement of the object. 
A Deep Dense Trajectory (DDT) model and a course learning scheme comprise the system.
Compared with other models \cite{DBLP:conf/eccv/ZhaoGRVMT18, DBLP:conf/eccv/GaoFG18} that rely on visual appearance cues, the system that relies on motion cues has a better performance on the separation of instrument sound sources.

\subsection{Audiovisual Generation}
Audio and visual generation are two active directions in the generation field.
Some studies \cite{DBLP:conf/cvpr/OwensIMTAF16, DBLP:conf/eccv/ChenZFWBN18, xie2007coupled, fang2020identity, DBLP:conf/cvpr/Hu0LN019} explore the relationship between audio and visual modalities to generate sound for vision, and vice versa.
Owens \emph{et al.} \cite{DBLP:conf/cvpr/OwensIMTAF16} synthesized sounds for silent videos of individuals beating and scraping items with drumsticks.
The generated sounds were realistic enough to even deceive subjects. 
Zhou \emph{et al.} \cite{DBLP:conf/cvpr/ZhouWFBB18} developed a video-to-sound model that directly predicts the waveform of sound from the input video.
The model is designed to generate sounds from videos in the wild rather than the laboratory setting as in \cite{DBLP:conf/cvpr/OwensIMTAF16}.
Chen \emph{et al.} \cite{DBLP:journals/tip/ChenZTXHG20} proposed a REGNET model that can generate time-aligned sounds for silent videos.
Conversely, sound can also guide the generation of vision, such as literature \cite{DBLP:conf/mm/ChenSDX17, DBLP:conf/cvpr/Hu0LN019} use the generative adversarial network to generate the corresponding visual image from the sound. 
Eskimez \emph{et al.} \cite{eskimez2021speech} took emotional label, speech, and a single face image as input to the model, and output a video of a talking face with audiovisual synchronization and emotional expression. 
Shlizerman \emph{et al.} \cite{shlizerman2018audio} accomplished sound-to-skeletal motion prediction by learning the relationship between hand motion and instrument performance.
The predicted skeleton information is applied to the 3D model to generate a performance video of the musical instrument.
Chen \emph{et al.} \cite{chen2019hierarchical} proposed to convert sound into facial landmarks and generate corresponding videos based on the landmarks. 
Compared to methods that directly learn audiovisual mappings, this method avoids fitting to sound-independent content.

\subsection{Binaural Stereo Audio Generation}
Binaural stereo audio generation refers to converting mono audio into binaural stereo audio using guidance from visual modalities or other spatial knowledge.
Morgado \emph{et al.} \cite{DBLP:conf/nips/MorgadoVLW18} proposed a model that utilizes mono audio and 360° video to generate and locate spatial audio.
Gao \emph{et al.} \cite{DBLP:conf/cvpr/GaoG19} used professional equipment to record the first video dataset containing binaural stereo audio indoors, and then combined the visual information with an encoder-decoder model to successfully convert mono audio into binaural stereo audio. 
Next, some studies \cite{DBLP:conf/eccv/ZhouXLWL20, DBLP:conf/cvpr/YangRS20} combine binaural stereo audio generation with a certain task to improve the performance.
Zhou \emph{et al.} \cite{DBLP:conf/eccv/ZhouXLWL20} observed that stereo audio generation and sound source separation have similar goals. They integrated these two tasks into a model to achieve sound source separation while improving the performance of stereo audio generation.
Yang \emph{et al.} \cite{DBLP:conf/cvpr/YangRS20} first learned the stereo channel-visual correspondence through a pretext task, and then used the learned embedding features to improve the performance of the model.
Compared to the 2D visual information used in previous methods, Chatziioannou \emph{et al.} \cite{lluis2022points2sound} proposed Point2Sound, which uses visual information in 3D point cloud format to guide the model to generate binaural stereo audio in the waveform domain. 
To alleviate the difficulty of recording visual-stereo pairs, Xu \emph{et al.} \cite{DBLP:conf/cvpr/XuZ0D0L21} constructed pseudo visual-stereo pairs by using mono audio and then used it to train mono-to-binaural network without real binaural recordings. 

There is a certain similarity between sound generation for silent video and binaural stereo audio generation, \emph{i.e.}, they are both generated by inference based on visual modalities.
The sound generated for silent video needs to be temporally and semantically aligned with the vision. 
However, binaural stereo audio generation pays more attention to the spatiality of the generated stereo audio.
For this reason, we concentrate more on the spatial analysis of binaural stereo audio, which motivates us to propose a spatial perception metric. 
This spatial evaluation of binaural stereo audio has not been considered in previous work.
Besides, different from these works, we introduce a shared visual-guided generative adversarial learning approach for stereo audio generation tasks and show that it can explore better vision-guided representations and have more spatially realistic generation effects.

\begin{figure*}[t]
  \centering
   \includegraphics[width=1\linewidth]{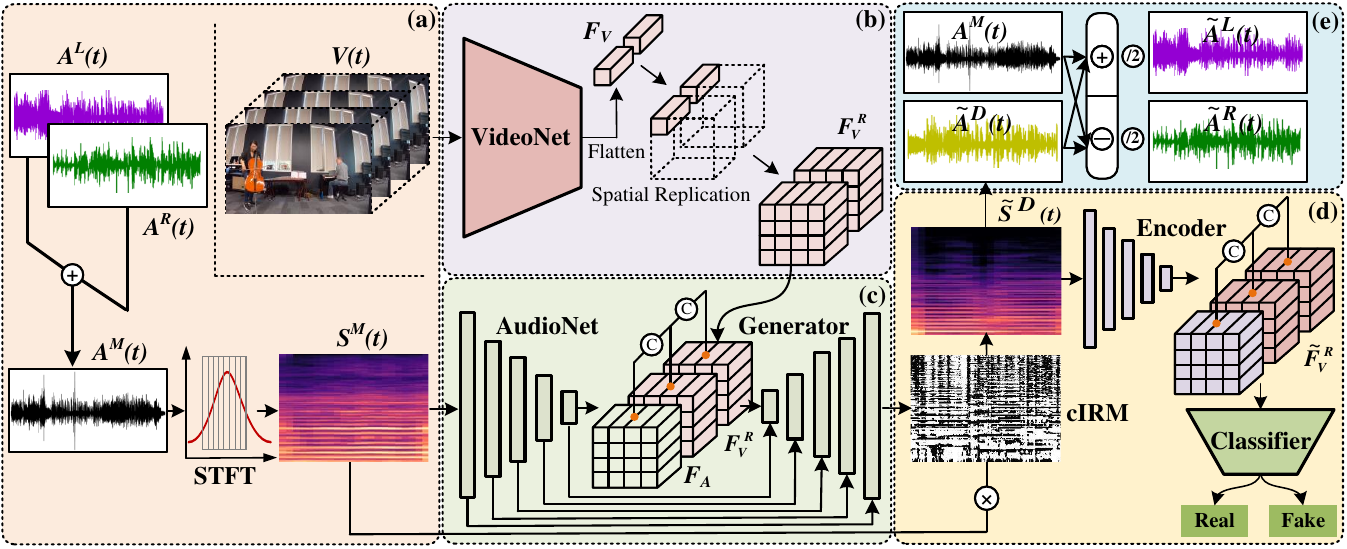}

   \caption{The pipeline of SAGM.
   The mixed spectrogram is produced by performing a short-time Fourier transform on the mixed audio.
   Then, the image sequence is fed into VideoNet to obtain spatio-temporal visual features, while the mixed spectrogram is fed into AudioNet to obtain audio features.  
   The fused audiovisual features are fed into the generator, which generates a cIRM to mask the mixed spectrogram.
   The discriminator takes the difference spectrogram and the updated visual features as input, and outputs whether the stereo audio is real or fake.
   Finally, stereo audio can be obtained by calculating between differential audio and mixed audio.
   The symbol {\scriptsize{\textcircled{\scriptsize{+}}}\scriptsize, \scriptsize{\textcircled{\scriptsize{--}}}\scriptsize, \scriptsize{\textcircled{\scriptsize{×}}}\scriptsize, \scriptsize{\textcircled{\tiny{/}\tiny{2}}}\scriptsize, \scriptsize{\textcircled{\tiny{C}}}\scriptsize} indicates element-wise sum, element-wise difference, element-wise complex product, element-wise halving, channel-wise concatenation, respectively.
   }
   \label{fig:twocol}
\end{figure*}

\section{Method}
A visually guided generative adversarial model is proposed, which generates binaural stereo audio from mono audio.
The SAGM approach is thoroughly described in this section.
First, the formulation of SAGM is introduced, which includes generator, discriminator, and loss. 
Then, the overall structure of SAGM is introduced, which consists of four components: {VideoNet}, {AudioNet}, binaural generator and discriminator.
In practice, other models can also be applied as components of SAGM as long as they achieve a similar power.

\subsection{SAGM Formulation}
The data used in our model is the video with binaural stereo audio. 
$V(t)$ is utilized to denote video, while $A^L(t)$ and $A^R(t)$ are utilized to denote the left and right channels of stereo audio, respectively. 
Next, we will introduce SAGM formulation in the order of generator formulation, discriminator formulation, and loss formulation. 

{\bf Generator formulation} gives the process of data preprocessing, complex ideal ratio mask generation, and data post-processing.
Mono audio can be obtained by:
\begin{equation}
  A^M(t) = A^L(t) + A^R(t).
  \label{eq:1}
\end{equation}
The mixing of left and right channel audio eliminates spatial effects in binaural stereo audio.
Compared to forecasting the left and right channels independently, the difference between channels can be more easily associated with visual information.
The difference between channels can be obtained by:
\begin{equation}
  A^D(t) = A^L(t) - A^R(t).
  \label{eq:2}
\end{equation}
The original waveform is converted into a spectrogram using the short-time Fourier transform (STFT) \cite{DBLP:conf/icassp/GriffinL83}:
\begin{equation}
  S^x(t) = {\rm STFT}(A^x(t)),
  \label{eq:3}
\end{equation}
where $x$ represents a type of audio.
For example, for mono audio, we use $S^M(t)$ to denote it.
So far, the generator formulation of the proposed SAGM can be written as: 
\begin{equation}
  \tilde{S}^D(t) = G(S^M(t), V(t)).
  \label{eq:4}
\end{equation} 

Recently, mask prediction has proven to be effective in singing voice or speech separation \cite{DBLP:conf/icassp/ZhangLW21, DBLP:conf/mwscas/HasannezhadOZC20} and stereo audio generation \cite{DBLP:conf/nips/MorgadoVLW18, DBLP:conf/eccv/ZhouXLWL20}.
Consequently, our objective is to predict a complex ideal ratio mask (cIRM) for the monophonic spectrogram to generate spectrogram of different channels. 
Note that because the result of the short-time Fourier transform of the sound wave is a complex result, $S^M(t)$ can be rewritten as 
\begin{equation}
  S^M(t) = S^M_{r}(t) + iS^M_{i}(t).
  \label{eq:5}
\end{equation}
The cIRM is also a complex from, it can be defined as
\begin{equation}
  M(t) = M_r(t) + iM_i(t),
  \label{eq:6}
\end{equation} 
where the subscripts $r$ and $i$ denote the real and imaginary components, respectively. 
Then,
\begin{equation}
  \tilde{S}^D(t) = S^M(t) * M(t),
  \label{eq:7}
\end{equation} 
where * denotes the complex multiplication rule. 
According to Eq. (7), Eq. (4) can be reformulated as
\begin{equation}
  S^M(t) * M(t) = G(S^M(t), V(t)).
  \label{eq:8}
\end{equation} 
Since $S^M(t)$ is known, the objective of SAGM is changed from the generation of channel difference spectrogram to the generation of cIRM. 

The spectrogram can be converted into a waveform signal using the inverse short-time Fourier transform (ISTFT) \cite{DBLP:conf/icassp/GriffinL83}:
\begin{equation}
  \tilde{A}^D(t) = {\rm ISTFT}(\tilde{S}^D(t)).
  \label{eq:9}
\end{equation}
Finally, the generated left and right channel audio can be calculated using the following equation: 
\begin{equation}
  \tilde{A}^L(t) = (A^M(t) + \tilde{A}^D(t))/2,
  \label{eq:10}
\end{equation} 
and
\begin{equation}
  \tilde{A}^R(t) = (A^M(t) - \tilde{A}^D(t))/2.
  \label{eq:11}
\end{equation}

{\bf Discriminator formulation} is the process of visually guiding the work of the discriminator.
Our idea is that the visual modality can guide the generator to generate stereo audio, and at the same time, the visual modality can also guide the discriminator to distinguish the true and false of the stereo audio. 
In this process, the discriminator is like an audience to discriminate the match between vision and sound.
Accordingly, we take the visual modality and the channel difference spectrogram together as the input of the discriminator:
\begin{equation}
  D(\tilde{S}^D(t), V(t))=0,
  \label{eq:12}
\end{equation}
and
 \begin{equation}
  D(S^D(t), V(t))=1.
  \label{eq:13}
\end{equation}

{\bf Loss formulation} is the process of generative adversarial learning.
SAGM can be optimized via the following min-max objective:
\begin{equation}
  G^* = {\rm arg}\ \underset{G}{\min}\ \underset{D}{\max}\ \mathcal{L}_{GAN}(G,D)
  \label{eq:14}
\end{equation}
where the adversarial loss $\mathcal{L}_{GAN}(G,D)$ is derived as:
\begin{equation}
\begin{aligned}
&\mathcal{L}_{GAN}(G,D) = E_{S^D(t) \sim p_{data}(S^D(t))}[{\rm log}D(S^D(t), V(t))]  \\&+ E_{S^M(t) \sim p_{data}(S^M(t))}[{\rm log}(1-D(G(S^M(t), V(t))))]
\end{aligned}
  \label{eq:15}
\end{equation}

During training under visual guidance, the discriminator $D$ is trained to reveal the differences between synthesized stereo audios and real stereo audios while the objective of the generator $G$ is to produce space-realistic results that can fool the discriminator $D$.

\begin{figure}[t]
  \centering
   \includegraphics[width=0.8\linewidth]{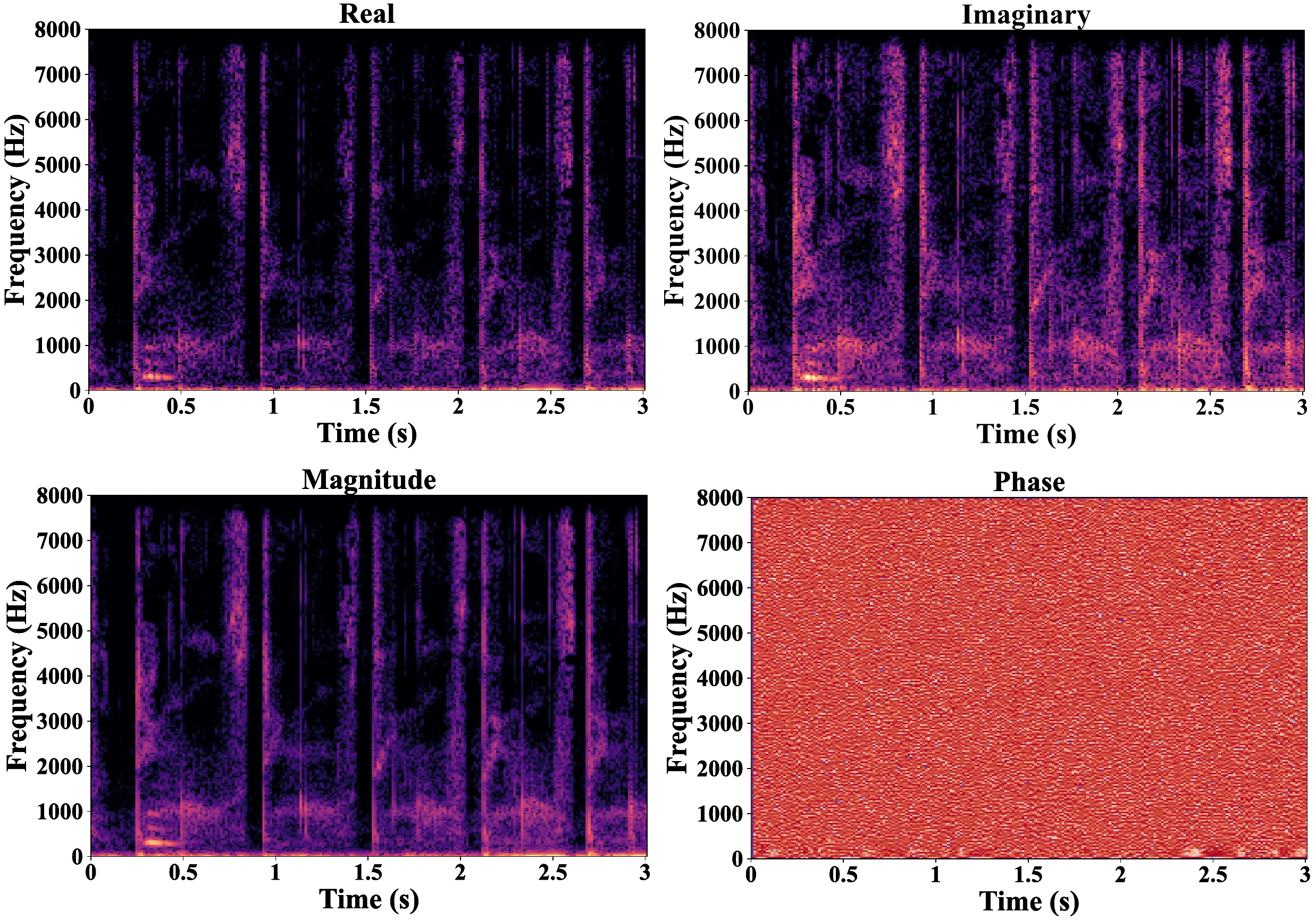}

   \caption{Examples of spectrograms.
Real spectrogram (top-left), imaginary spectrogram (top-right) and magnitude spectrogram (bottom-left) have a salient time-frequency structure.
Phase spectrogram (bottom-right) shows an inconspicuous time-frequency structure, just like noise.}
   \label{fig:threecol}
\end{figure}

\subsection{Overall Architecture}
SAGM takes mono audio $A^M(t)$ and video $V(t)$ as input, and outputs left channel audio $\tilde{A}^L(t)$ and right channel audio $\tilde{A}^R(t)$. 
The overall procedure can be illustrated in Fig. \ref{fig:twocol}.
Fig. \ref{fig:twocol}(a) shows the data preprocessing process of SAGM.
As shown in Fig. \ref{fig:twocol}(b), VideoNet is used to extract spatio-temporal visual features for the stereo generator and discriminator. 
Since binaural stereo audio generation is related to the spatial position and motion state of the object, we argue that spatio-temporal visual features provide greater guidance gain than spatial features, especially when the sounding object is in motion.
We utilize 3D convolutional neural network C3D \cite{DBLP:conf/iccv/TranBFTP15} to extract spatio-temporal visual features $F_V$. 
The fully connected layer of C3D is removed and the time downsampling is reduced from 16 to 4 to meet the proposed model. 
The size of final visual feature extracted by C3D is ($T/4$) × ($H/32$) × ($W/32$) × $C$, where $T$, ($H$, $W$), $C$ denotes the temporal dimensions, image size and channel dimensions.
In Fig. \ref{fig:twocol}(c), AudioNet with 5 convolutional layers accepts the mixed audio spectrogram as input.
Initially, the mixed audio spectrogram is in the form of complex numbers. 
Subsequently, the imaginary and real components of the spectrogram are concatenated in an extra dimension to obtain a new spectrogram form. 
The intuitive idea is to generate stereo audio in the phase and amplitude of the audio, instead of concatenating the real and imaginary parts together. 
However, we found through experiments that this approach is difficult.  
Because the phase spectrum shows a fragile time-frequency structure (see Fig. \ref{fig:threecol}). 
AudioNet is utilized to extract mono audio features $F_A$ that contain mixed semantic information but no spatial information. 
The dimension of the extracted mono audio feature is  ($H/32$) × ($W/32$) × $C$, where ($H$, $W$), $C$ denotes spectrogram size and channel dimensions. 
To provide pixel-wise spatio-temporal visual guidance information to audio features, spatial visual features are flattened in the channel dimension.
Next, the flattened visual features are spatially copied to obtain new visual features $F_V^R$.
The new visual features have the same spatial size as $F_A$.
Finally, $F_V^R$ is channel-wise concatenated with $F_A$, obtaining the final fused audiovisual features.

Binaural generator takes the fused audiovisual features as input. 
It is composed of 5 up-sampling convolutional layers.
The hierarchical feature size of the stereo generator is the same as that of AudioNet, so the features can be concatenated by layer jumps.
The generator's upsampling size is 32.
The output of the generator is $M(t)$, which represents the cIRM after Tanh activation.
As shown in Fig. \ref{fig:twocol}(d), the function of $M(t)$ is to predict the spectrogram $\tilde{S}^D(t)$ of the differential signal by masking the spectrogram of the mixed mono audio.
The goal of the discriminator is to distinguish the realism of binaural stereo audio.
A straightforward approach is to feed the binaural stereo audio into the discriminator alone to predict authenticity.
This approach can also make the discriminator work, nevertheless, its interpretability is limited.
Our idea is that the discriminator can identify the authenticity of stereo audio by combining visual information and stereo audio information like a stereo video viewer.
We introduce visual modality into the discriminator to assist the discriminator to distinguish the authenticity of stereo audio, just as we introduce visual modality to guide the generator to generate stereo audio.
As mentioned earlier, the differential signal of stereo audio is easier to associate with visual information.
Therefore, we utilize the channel difference spectrogram as the input of the discriminator, instead of the spectrogram of the entire stereo audio. 
The spatio-temporal visual features input to the discriminator are the updated visual features $\tilde{F}^R_V$ in the generator.
The updated visual features of the discriminator are also re-used as the visual input of the generator.
The encoder of discriminator adopts a network similar to AudioNet to extract audio features from the spectrogram of the channel difference. 
The dimension of the extracted differential audio feature is  ($H/32$) × ($W/32$) × $C$, where ($H$, $W$), $C$ denotes spectrogram size and channel dimensions. 
The visual features and differential audio features of the discriminator are fused in the same way as the generator.
Finally, the fused audiovisual features are input into two convolutional layers and an average pooling layer to output the classification results. 
We optimize the discriminator using binary cross entropy loss. 
The L1 loss is used to train generator to minimize the distance between $\tilde{S}^D(t)$ and the ground-truth $S^D(t)$. 
In Fig. \ref{fig:twocol}(e), the generated difference audio $\tilde{A}^D(t)$ can be obtained by Eq. (9), and the generated stereo audio can be obtained by Eq. (10) and Eq. (11).  
In the generative inference phase, the generator receives mono audio and the associated visual modality as input and produces stereo audio.

\begin{figure*}[t]
  \centering
   \includegraphics[width=1\linewidth]{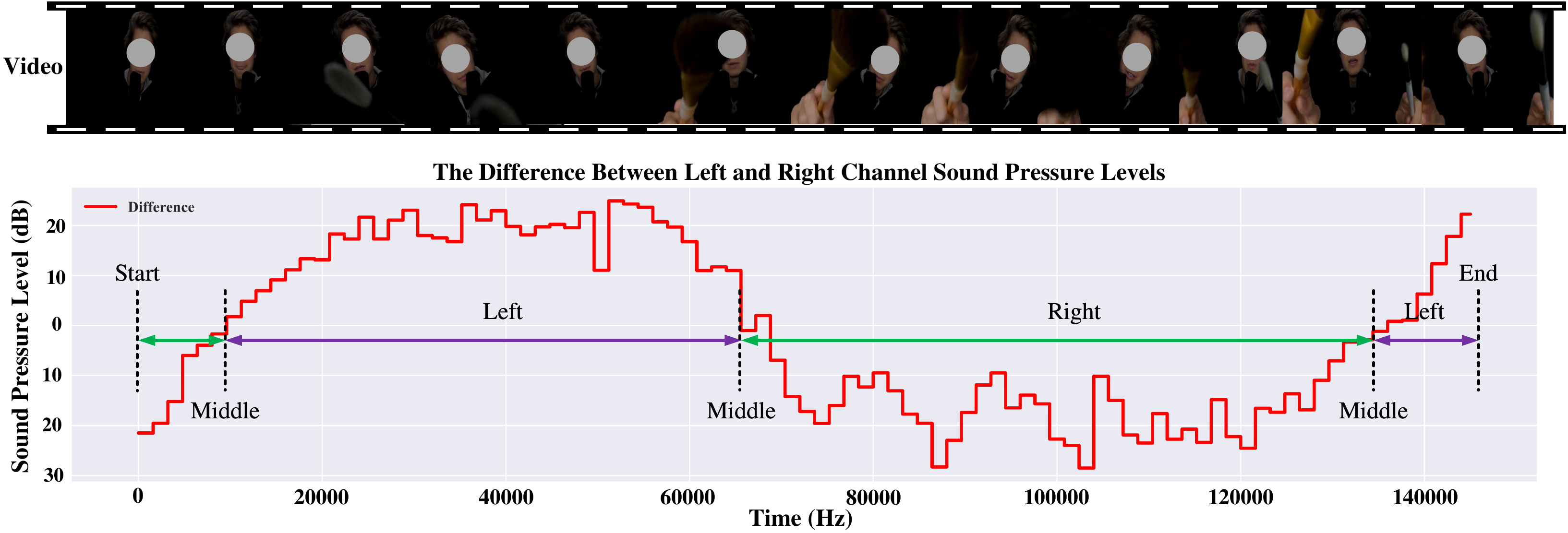}

   \caption{An example of the spatial perception variation curve of stereo audio. 
   Spatial perception is defined as the difference between the sound pressure levels of the left and right channels. 
   The curve above the 0 dB scale means that the stereo sound at the moment is biased towards the left channel, and on the contrary, it is biased towards the right channel. 
   }
   \label{fig:fourcol}
\end{figure*}

\section{Experiments}
\subsection{Datasets}
To demonstrate the effectiveness and superiority of the proposed SAGM, we evaluated it on two benchmark datasets:

{\bf FAIR-Play} \cite{DBLP:conf/cvpr/GaoG19} dataset is the first to be recorded in a laboratory with professional equipment, and the data is video with binaural audio.
The dataset is characterized by recording in a way that simulates binaural reception of sound. 
This dataset consists of 1871 10-second clips of musical performances, totaling 5.2 hours.
The dataset has 10 random segmentations. 
In each segmentation, the train set, validation set, and test set are divided into 1497/187/187, respectively.

{\bf YT-Music} dataset comprises 397 music performance videos of different durations.
These videos are collected by \cite{DBLP:conf/nips/MorgadoVLW18} on YouTube. 
This dataset is characterized by a great number of blended sound sources. 
The video and audio in the YT-Music dataset are 360° view \cite{li2018scene}. 
We follow the rules of \cite{DBLP:conf/cvpr/GaoG19} to convert first-order ambisonics audios into binaural audios.
\subsection{Baseline Methods}
Several methods are compared with the proposed method.
They are introduced as follows:

{\bf MONO2BINAURAL} \cite{DBLP:conf/cvpr/GaoG19} is the first baseline model on the task of binaural stereo audio generation. 
The model is based on an autoencoder structure and utilizes ResNet-18 pre-trained on ImageNet to extract spatial visual information. 

{\bf Sep-stereo} \cite{DBLP:conf/eccv/ZhouXLWL20} unifies stereo audio generation and sound source separation into a single framework.
The sound source separation branch and the binaural stereo audio generation branch share a backbone network.
In addition, an associative pyramid network (APNet) is added to the Sep-stereo model for better fusing visual and audio features.
The best generation performance of this method requires an additional sound source separation dataset. 

{\bf Main network} \cite{DBLP:conf/mir/ZhangS21} employs an audiovisual attention fusion module (AVAFM) to integrate visual features and audio features.  
MAFNet in the Main network fuses different levels of audio and visual features by stacking multiple AVAFMs. 
This method has better fusion features than MONO2BIN-
AURAL, Sep-stereo, and our method. 

{\bf Complete network} \cite{DBLP:conf/mir/ZhangS21} adds an iterative network based on the Main network to optimize stereo audio generation through iteration. 

\subsection{Implementation Details}
We randomly split a 1s segment from a 10s stereo video clip to train our model. 
For audio, we resample stereo audio at 16kHz.
For video, we extract 4 frames in 1s. 
STFT is computed using a Hann window of length 25ms, hop length of 15ms, and FFT size of 448.
The size of the spectrogram and image are 225×64 and 224×224 respectively. 
We use the Adam optimizer to train the model with a batch size of 24. 
The learning rate of VideoNet, AudioNet, binaural generator, and discriminator is set to 0.001. 
The training is terminated when the number of model training epochs reaches 1000.
During inference, we utilize a window with a hop size of 0.1s to slide over the mono audio to generate 10s stereo audio.
Our computational environment is an Intel (R) Xeon (R) E5-2680 v4 CPU, @ 2.40GHz, Ubuntu 18.04, and nvidia GeForce RTX 3090. 
The implementations of deep learning models are realized by PyTorch1.7.

\subsection{Evaluation Metrics}
Six evaluation metrics are adopted to compare the proposed SAGM with other advanced methods, including STFT Distance, Envelope (ENV) Distance, Wave L2, Multi-Resolution STFT (MRSTFT), Signal-to-Noise Ratio (SNR), and Sound Pressure Level (SPL) Distance.
The first two metrics are commonly used in binaural stereo audio generation.
In addition, we also evaluate three new metrics: Wave L2, MRSTFT and SNR.
The above five metrics only reveals the distance between the predicted signal and the real signal, and does not reflect the stereoscopic difference between them.
To this end, we propose a SPL Distance to evaluate the difference in sound spatial perception.

{\bf STFT Distance} measures binaural stereo audio on the spectrogram domain, which is the Euclidean distance between the predicted left and right channel spectrograms and their ground-truth:
\begin{equation}
  \mathcal{D}_S = {\rVert S^L(t) - \tilde{S}^L(t)\rVert}_2 + {\rVert S^R(t) - \tilde{S}^R(t)\rVert}_2.
  \label{eq:16}
\end{equation}

{\bf ENV Distance} measures binaural stereo audio on the raw waveform domain, which is the Euclidean distance between the envelope of the predicted waveform's left and right channels and its ground-truth:
\begin{equation}
  \mathcal{D}_E = {\rVert E[A^L(t)]- E[\tilde{A}^L(t)]\rVert}_2 + {\rVert E[A^R(t)]- E[\tilde{A}^R(t)]\rVert}_2,
  \label{eq:17}
\end{equation}
where $E[A(t)]$ denote the envelope of signal $A(t)$.
It can capture the perceptual similarity of the raw waveform well. 
{\bf Wave L2} is the mean squared error between the generated stereo audio and the real stereo audio.
{\bf MRSTFT} comprehensively considers the effects of spectral convergence, log magnitude loss and linear magnitude loss on multi-resolution spectrum loss \cite{leng2022binauralgrad}.
{\bf SNR} is the power ratio of the stereo audio signal to the noise.
{\bf SPL Distance} is the Euclidean distance between the SPL difference between the left and right channels of the real signal and the SPL difference between the left and right channels of the predicted signal.
The definition of SPL is as follows: 
\begin{equation}
  SPL(t) = 20 \times {\rm log_{10}}(\frac{{\rVert A(t) \rVert}_2}{p_{ref}}), 
  \label{eq:18}
\end{equation}
where $A(t)$ represents the original waveform, $p_{ref}$ = $2 \times 10^{-5}$. Then, the $SPL(t)$ of the left and right channels of the real signal and the predicted signal can be expressed as $SPL^L(t)$, $SPL^R(t)$, $\tilde{SPL}^L(t)$, $\tilde{SPL}^R(t)$, respectively. 
The spatial perception of a stereo audio is defined as the difference between the $SPL$ of the left channel and the right channel, where $\pm$ indicate the direction. 
Then, the spatial perception of the ground-truth stereo audio and the predicted stereo audio is
\begin{equation}
  SD_{SPL}(t) = SPL^L(t) - SPL^R(t), 
  \label{eq:19}
\end{equation}
and 
\begin{equation}
  \tilde{SD}_{SPL}(t) = \tilde{SPL}^L(t) - \tilde{SPL}^R(t). 
  \label{eq:20}
\end{equation}
Finally, SPL Distance can be calculated by:
\begin{equation}
  \mathcal{D}_{SPL}(t) = {\rVert SD_{SPL}(t) - \tilde{SD}_{SPL}(t) \rVert}_2.   
  \label{eq:21}
\end{equation}

In this paper, we utilize SPL Distance to visualize the spatial perception of binaural stereo audio for qualitative evaluation. 
Fig. \ref{fig:fourcol} is an example of we use of the $SD_{SPL}(t)$ to visualize the spatial perception of binaural stereo audio. 
The spatial trajectory of the speaker in the video is right$\rightarrow$left$\rightarrow$right$\rightarrow$left. 
It can be seen that the curve in the figure well reflects the variation in the spatial perception direction of binaural stereo audio.
The spatial perception metric of stereo audio reflects the direction and magnitude of stereo spatial perception, expressed as: 

\begin{equation}
  Dir(t) = \begin{cases}
  \rm{Right}, &  SD_{SPL}(t) < 0 \\
  \rm{Middle}, & SD_{SPL}(t) = 0 \\
  \rm{Left}, & SD_{SPL}(t) > 0 \\ 
             \end{cases}
  \label{eq:22}
\end{equation}
and 
\begin{equation}
  Mag(t) = \mid SD_{SPL}(t) \mid.
  \label{eq:23}
\end{equation}

\section{Quantitative and Qualitative Results}
We perform diverse and comprehensive experiments to compare the proposed method with other methods, including quantitative experiments, qualitative experiments, parametric experiments, and user studies.
In ablation experiments, we verify the effectiveness of visual guidance, visual pre-training, visual spatio-temporal information, and visual information sharing.

\begin{table}
\begin{center}
\caption{Quantitative results of SAGM on FAIR-Play dataset.}
\label{tab1}
\resizebox{13cm}{17mm}{
\begin{tabular}{| c | c | c | c | c | c |}
\hline
Method & STFT$\downarrow$ & ENV$\downarrow$  & Wave L2 ($\times10^3$) $\downarrow$ & MRSTFT$\downarrow$ & SNR$\uparrow$\\
\hline
MONO2BINAURAL \cite{DBLP:conf/cvpr/GaoG19}& 0.959 & 0.141  &6.496 &0.940 &6.232\\
\hline
APNet \cite{DBLP:conf/eccv/ZhouXLWL20} & 0.889  & 0.136  &5.758 &0.944 &6.972\\
\hline
Sep-stereo \cite{DBLP:conf/eccv/ZhouXLWL20} & 0.879  & 0.135  &6.526 &0.962 &6.422\\
\hline
Main network \cite{DBLP:conf/mir/ZhangS21} & 0.867  & 0.135  &5.750 &0.950 &6.985\\
\hline 
Complete network \cite{DBLP:conf/mir/ZhangS21} & 0.856  & 0.134  &5.787 &0.948 &6.959\\
\hline 
SAGM (Ours) & \textbf{0.851} & \textbf{0.134}  &\textbf{5.684} &\textbf{0.914} &\textbf{7.044}\\
\hline 
\end{tabular}}
\end{center}
\end{table}

\subsection{Quantitative Results}
The results of 10 cross-validation experiments on the FAIR-Play dataset are shown in Table \ref{tab1}.
Compared with other methods, the proposed SAGM achieves the best performance on all metrics.
SAGM has a great improvement in performance compared to the MONO2BINAURAL baseline model (STFT: 0.108$\downarrow$, ENV: 0.007$\downarrow$, Wave L2: 0.812$\downarrow$, MRSTFT: 0.026$\downarrow$, SNR: 0.812$\uparrow$).
Sep-stereo combines the binaural stereo audio generation with the sound source separation, and the two tasks share an encoder-decoder structure.
Its best experimental results rely on additional sound source separation datasets.
Compared to Sep-stereo, our method focuses on learning more powerful visual guidance rather than the encoding-decoding capabilities of sound.
The proposed SAGM adopts a single-task approach to achieve better results than Sep-stereo using a multi-task method (STFT: 0.851 \emph{vs} 0.879, ENV: 0.134 \emph{vs} 0.135, Wave L2: 5.684 \emph{vs} 6.526, MRSTFT: 0.914 \emph{vs} 0.962, SNR: 7.044 \emph{vs} 6.422).
The proposed method also outperforms recently proposed multi-attention fusion method.
The advanced performance of Complete network benefits from its attention-based fusion method and iterative network structure. 
Since SAGM adopts a spatio-temporal visual-guided generative adversarial training paradigm, it can achieve better performance than Complete network only by directly concatenating visual and audio features (STFT: 0.851 \emph{vs} 0.856, Wave L2: 5.684 \emph{vs} 5.787, MRSTFT: 0.914 \emph{vs} 0.948, SNR: 7.044 \emph{vs} 6.959).
It is worth mentioning that the proposed SAGM can easily integrate multi-task and audiovisual feature fusion methods to further improve the performance of SAGM.
Thus, the proposed SAGM is a general generation framework, \emph{i.e.}, a binaural stereo audio generative adversarial approach under shared visual guidance.

\begin{table}
\begin{center}
\caption{Quantitative results of SAGM on YT-Music dataset.}
\label{tab2}
\resizebox{13cm}{17mm}{
\begin{tabular}{| c | c | c | c | c | c |}
\hline
Method & STFT$\downarrow$ & ENV$\downarrow$  & Wave L2 ($\times10^3$) $\downarrow$ & MRSTFT$\downarrow$ & SNR$\uparrow$\\
\hline
MONO2BINAURAL \cite{DBLP:conf/cvpr/GaoG19} & 1.346 & 0.179  &6.337 &0.938 &5.008\\
\hline
APNet \cite{DBLP:conf/eccv/ZhouXLWL20} & 1.070 & 0.148  &5.805 &0.936 &5.542\\
\hline
Sep-stereo \cite{DBLP:conf/eccv/ZhouXLWL20} & 1.051  & 0.145  &6.323 &1.108 &4.779\\
\hline
Main network \cite{DBLP:conf/mir/ZhangS21} & 1.036  & 0.144  &5.944 &0.936 &5.573\\
\hline 
Complete network \cite{DBLP:conf/mir/ZhangS21} & 1.023  & 0.142  &6.313 &1.053 &4.873\\
\hline 
SAGM (Ours) &\textbf{0.875} & \textbf{0.126}  &\textbf{5.792} &\textbf{0.916} &\textbf{5.601}\\
\hline 
\end{tabular}}
\end{center}
\end{table}

Compared with the FAIR-Play dataset, the YT-Music dataset is collected from diverse indoor and outdoor environments.
Therefore, the dataset contains numerous sound sources and interfering noises, which makes this dataset more challenging.
Table \ref{tab2} demonstrates the experimental results of the YT-Music dataset under split1 segmentation.
The experimental results reveal that the proposed SAGM also outperforms other methods on challenging dataset.
Compared with the second best method, the proposed SAGM improves the STFT, ENV, Wave L2, MRSTFT, and SNR metrics by 0.148$\downarrow$, 0.016$\downarrow$, 0.521$\downarrow$, 0.137$\downarrow$, and 0.728$\uparrow$, respectively.
This indicates that the proposed method has good application potential.

\subsection{Parameters and Ablation Experiment Results} 
We implement parameter experiments to verify the impact of different parameters on the binaural stereo audio generation model.
Table \ref{tab3} is the result of parameter experiment.
We select 2 methods and 5 parameters for comparison.
$FPS$, $I$, $S$, $L$, and $Hop$ respectively represent the sampling rate of the video, the resolution of the image, the resolution of the spectrogram, the length of the audio, and the length of the sliding window for inference. 
In the parameter experiment of the MONO2BINAURAL model, it can be seen that reducing the resolution of the image and spectrogram leads to a decrease in generation performance (STFT: 0.009$\uparrow$, ENV: 0.001$\uparrow$).
However, the performance of the proposed SAGM under low-resolution input is ahead of the MONO2BINAURAL model under high-resolution input (STFT: 0.868 \emph{vs} 0.945, ENV: 0.135 \emph{vs} 0.141).
It can be seen from SAGM that the length of the sliding window for inference has a significant impact on the generation performance (STFT: 0.868 \emph{vs} 0.965, ENV: 0.135 \emph{vs} 0.149). 
The shorter the length of the sliding window, the better the model generation performance, but this will increase the number of inferences of the model.

\begin{table*}[!t]
\begin{center}
\caption{The result of parameter experiment. All parameter experiments are implemented on split1 of FAIR-Play dataset.}
\label{tab3}
\resizebox{\textwidth}{15mm}{
\begin{tabular}{| c | c | c | c | c | c | c | c |}
\hline
Method & FPS & I & S & L & Hop & STFT & ENV\\
\hline
MONO2BINAURAL \cite{DBLP:conf/cvpr/GaoG19} & 4 & 448 × 224 & 257 × 64 & 0.63 & 0.05 & 0.945  & 0.141 \\
\hline
MONO2BINAURAL \cite{DBLP:conf/cvpr/GaoG19} & 4  & 224 × 224 &225 × 64 & 1 & 0.05 & 0.954 & 0.142 \\
\hline
SAGM (Ours) & 4 & 224 × 224 &225 × 64 & 1 & 1 & 0.965 & 0.149 \\
\hline
SAGM (Ours) & 4 & 224 × 224 &225 × 64 & 1 & 0.1 & 0.869 & 0.135 \\
\hline
SAGM (Ours) & 4 & 224 × 224 &225 × 64 & 1 & 0.05 & 0.868 & 0.135 \\
\hline 
\end{tabular}}
\end{center}
\end{table*}

\begin{table}
\begin{center}
\caption{Ablation results of SAGM on FAIR-play dataset.}
\label{tab4}
\resizebox{8.5cm}{19.5mm}{
\begin{tabular}{| c | c | c |}
\hline
Method & STFT & ENV\\
\hline
SAGM (w/o V) & 0.9973 & 0.1450 \\
\hline
SAGM (w/o shared) & 0.8769 & 0.1373 \\
\hline
SAGM (ResNet w/o pretrained) & 0.8770 & 0.1369 \\
\hline
SAGM (ResNet) & 0.8708 & 0.1363 \\
\hline
SAGM (C3D w/o pretrained) & 0.8775 & 0.1367 \\
\hline
SAGM ($D$ w/o V) & 0.8733  & 0.1360 \\
\hline
SAGM (C3D) & 0.8687 & 0.1354 \\
\hline
\end{tabular}}
\end{center}
\end{table}

\begin{figure*}[t]
  \centering
   \includegraphics[width=1\linewidth]{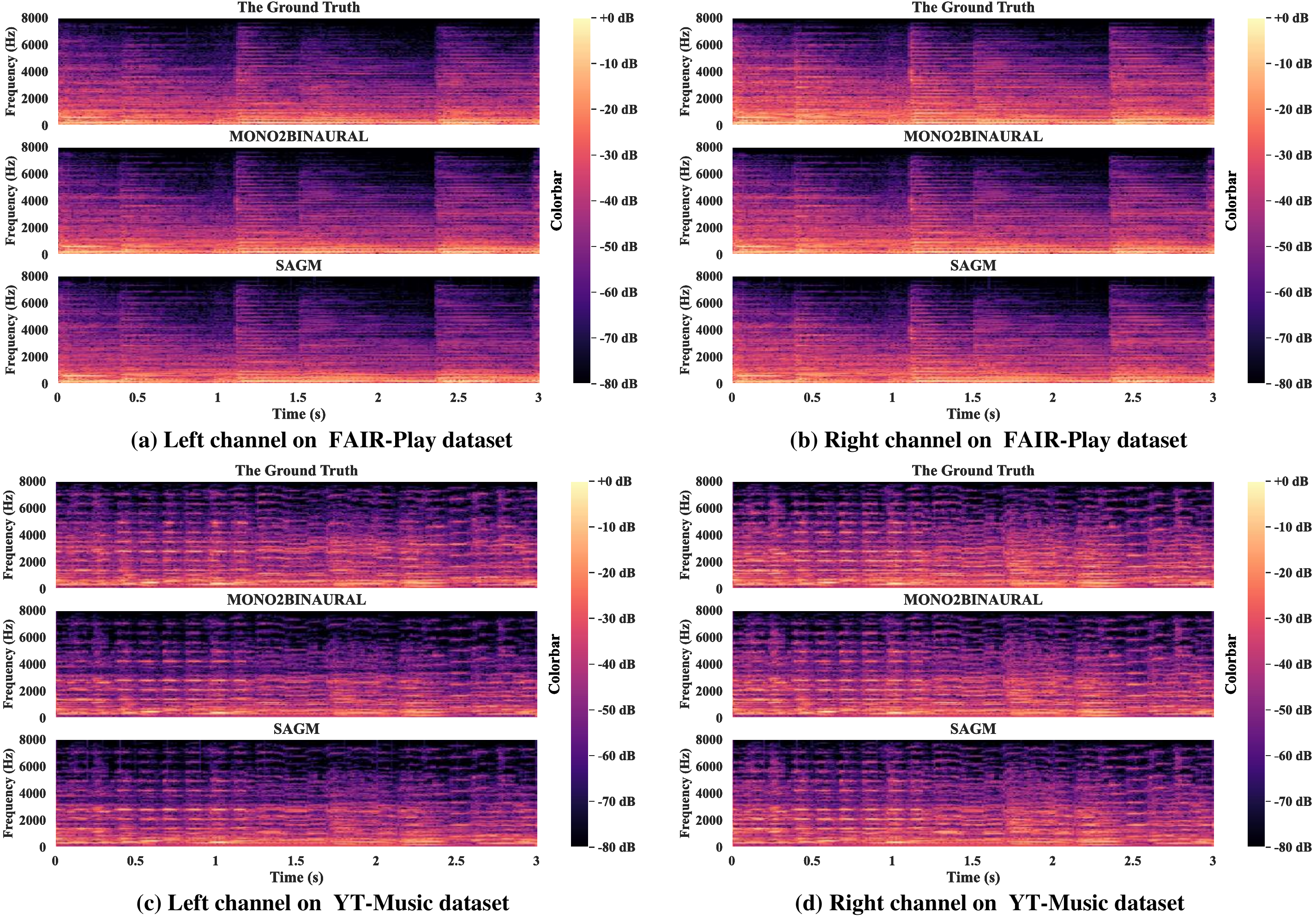}

   \caption{Stereo audio frequency spectrum visualization results.
(a) is the spectrogram of the stereo left channel, while (b) is the spectrogram of the stereo right channel on FAIR-Play dataset. 
(c) is the spectrogram of the stereo left channel, while (d) is the spectrogram of the stereo right channel on YT-Music dataset. 
In order to enhance the visualization of the spectrum, we convert the amplitude spectrum to a dB scale spectrum.}
   \label{fig:fivecol}
\end{figure*}

In Table \ref{tab4}, we demonstrate the importance of visual information for binaural stereo audio generation and validate the effectiveness of alternating adversarial learning under visual sharing. 
SAGM (w/o V) removes VideoNet, \emph{i.e.}, the model is not visually guided.
SAGM (ResNet) refers to SAGM using ResNet-18 pre-trained on ImageNet as a feature extractor to extract visual spatial features.
SAGM (C3D) is SAGM using C3D pre-trained on Sports-1M as a feature extractor to extract visual spatio-temporal features.
The results of ablation experiments show that visual information can provide guidance for binaural stereo audio generation (STFT: 0.8687 \emph{vs} 0.9973, ENV: 0.1354 \emph{vs} 0.1450).
Compared with visual spatial features, visual spatio-temporal features can provide superior visual guidance information for binaural stereo audio generation (STFT: 0.8687 \emph{vs} 0.8708, ENV: 0.1354 \emph{vs} 0.1363). 
The pre-trained VideoNet can further improve model performance (C3D $||$ STFT: 0.8687 \emph{vs} 0.8775, ENV: 0.1354 \emph{vs} 0.1367) (ResNet $||$ STFT: 0.8708 \emph{vs} 0.8770, ENV: 0.1363 \emph{vs} 0.1369).
Visual sharing enables guidance information to be transmitted between the generator and the discriminator to learn bidirectional complementary visual guidance.
In SAGM (w/o shared), we remove the visual sharing setting between the generator and the discriminator, which leads to a drop in model performance (STFT: 0.0082$\uparrow$, ENV: 0.0019$\uparrow$). 
This demonstrates the effectiveness of visual sharing in binaural stereo audio generation.
In addition, we remove the visual guidance information of the discriminator to verify its impact on the performance of SAGM. 
The discriminator in SAGM (D w/o V) has no visual guidance, compared to the discriminator in SAGM (C3D). 
It can be seen that the performance of SAGM (D w/o V) is degenerate (STFT: 0.0046$\uparrow$, ENV: 0.0006$\uparrow$).
The reason is that the discriminator with visual guidance has stronger discriminating ability, and the adversarial learning between the discriminator and the generator leads to a better performance of the generator.
This indirectly demonstrates the effectiveness of the visually guided generative adversarial network in learning bidirectional complementary visual guidance.

 \begin{figure*}[t]
  \centering
   \includegraphics[width=0.9\linewidth]{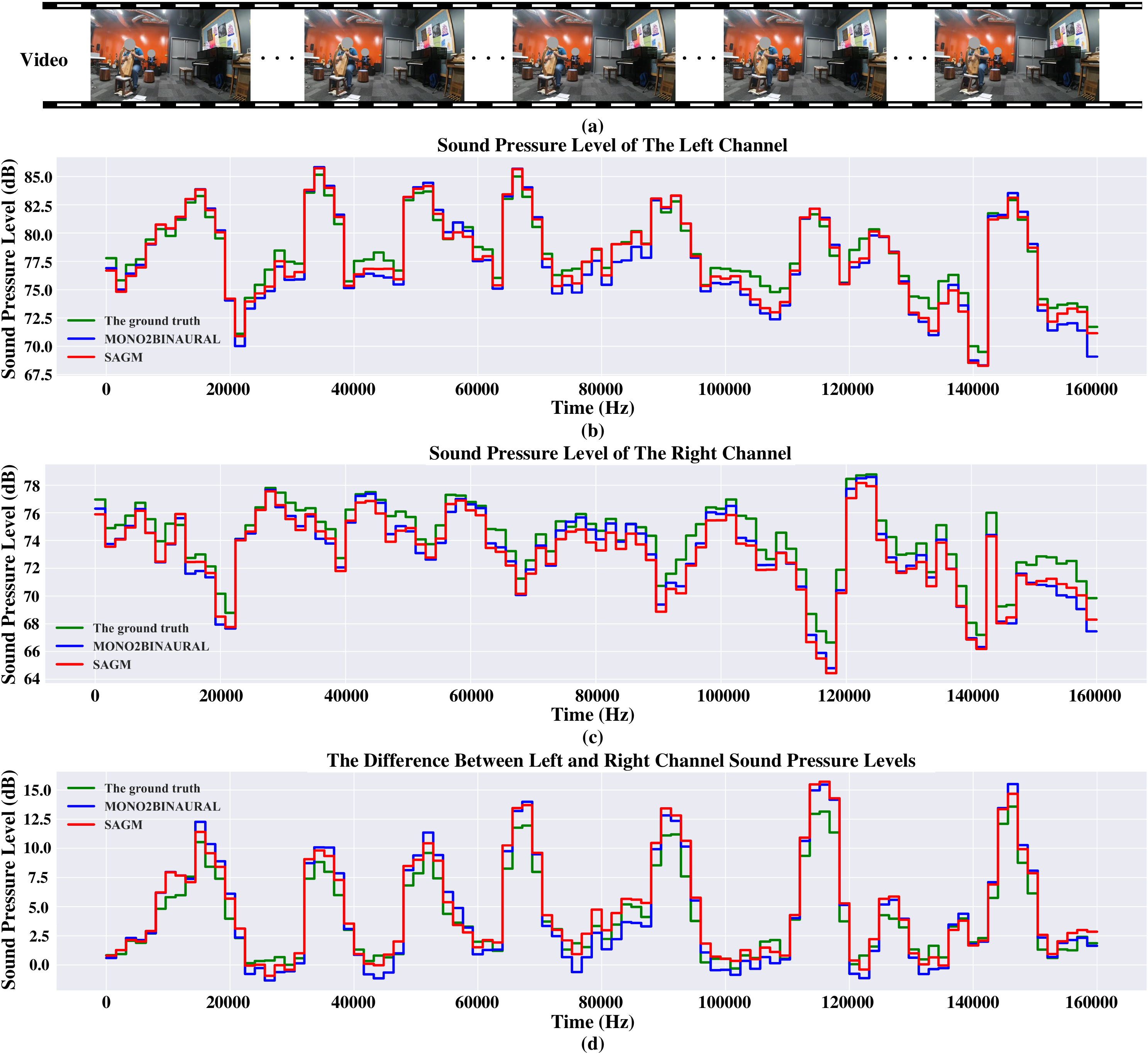}

   \caption{Sound pressure level characterization on the FAIR-Play dataset.
   (a) is the video. 
   (b), (c) and (d) are the sound pressure level curve of the left channel, the sound pressure level curve of the right channel, and the difference sound pressure level curve between them. 
   }
   \label{fig:sixcol}
\end{figure*}

\subsection{Qualitative Results}
We implement qualitative experiments on the FAIR-Play dataset and YT-Music dataset, and compare our method with MONO2BINAURAL model.
In Fig. \ref{fig:fivecol}, we visualize the spectrogram of the left and right channels of binaural stereo audio.
The upper image ((a), (b)) is the experimental result on the FAIR-Play dataset, while the lower image ((c), (d)) is the experimental result on the YT-Music dataset. 
Fig. \ref{fig:fivecol} shows that both our method and the baseline model generate good spectrograms of the left and right channels.
The generated binaural stereo audio and the ground-truth have a similar time-frequency structure, which reflects the generation performance of the model to a certain extent. 
However, it is not sufficient to present the qualitative results of the model using spectrograms.
The first reason is that the spectrograms of the same channel are quite similar, making it difficult to compare experimental results between diverse methods.
Another reason is that spectrograms fail to reflect the spatial perception of binaural stereo audio.
In Fig. \ref{fig:fivecol}, it is challenging to clearly observe the spatial effect from the comparison of the spectrograms of the left and right channels.
The goal of the binaural stereo audio generation task is to generate realistic audio with a sense of space.
Therefore, it is extremely necessary to evaluate the spatial perception of binaural stereo audio.
The proposed SPL Distance metric is well capable of assessing the spatial perception of binaural stereo audio.

\begin{figure*}[t]
  \centering
   \includegraphics[width=0.9\linewidth]{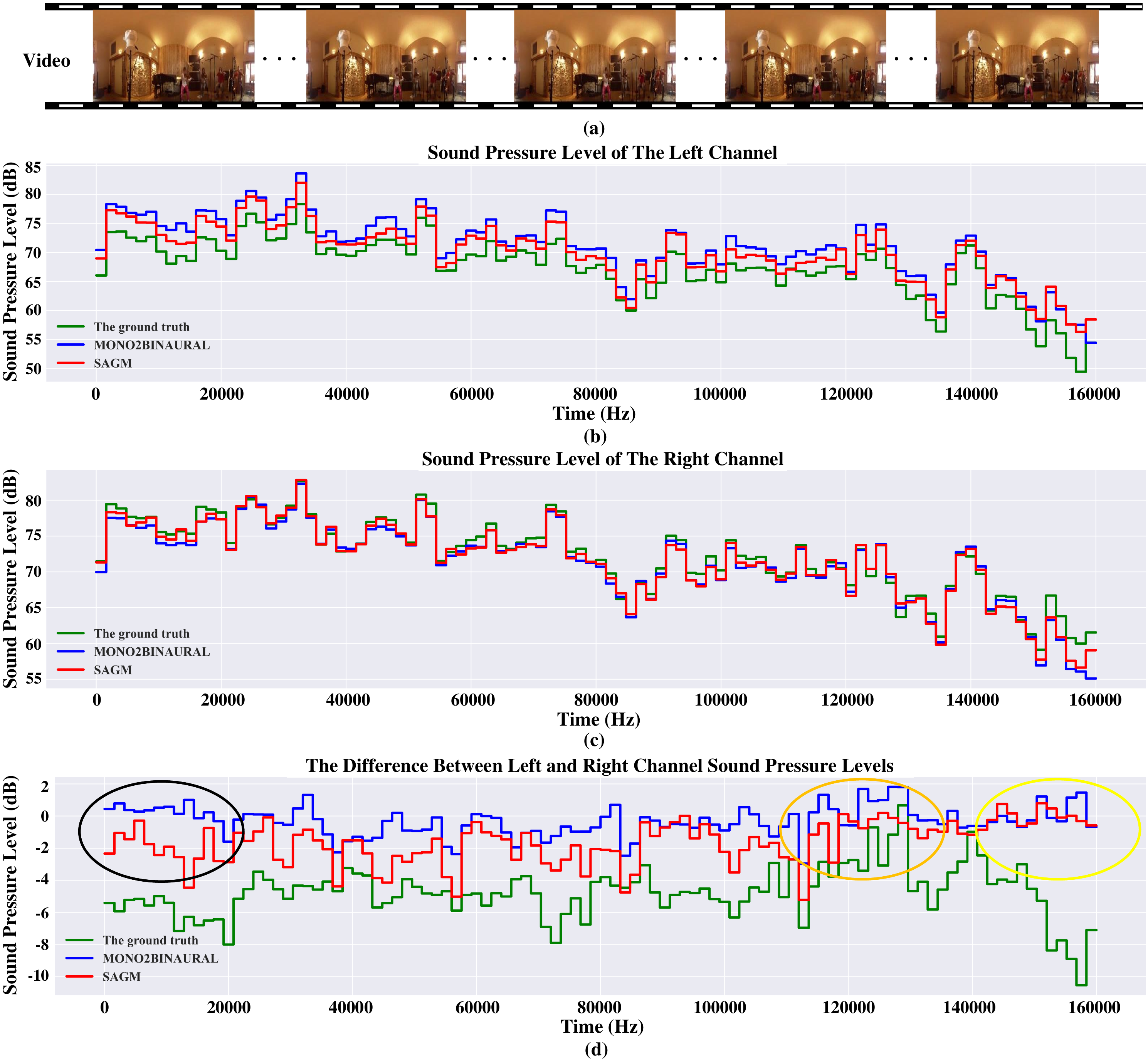}
   \caption{Sound pressure level characterization on the YT-Music dataset.
   (a) is the video. 
   (b), (c) and (d) are the sound pressure level curve of the left channel, the sound pressure level curve of the right channel, and the difference sound pressure level curve between them. 
   }
   \label{fig:sevencol}
\end{figure*}

The experimental results of spatial perception evaluation on FAIR-Play dataset are shown in Fig. \ref{fig:sixcol}. 
It shows the sound pressure levels of the left (b) and right (c) channels of binaural stereo audio and their difference curve (d).
We exhibit this figure because we observe the relationship between the sound pressure level of the channel and the spatial perception of the audio.
Spatial perception is displayed on the side with higher sound pressure levels when the sound pressure levels of the channels are different.
This phenomenon can be seen from Fig. \ref{fig:sixcol}(d).
The drum and harp are played on the left in Fig. \ref{fig:sixcol}(a). 
The spatial perception of the audio is shown on the left. 
The sound pressure level difference curve of the left and right channels is always above the 0 dB scale. 
It can be seen from the figure that both SAGM and MONO2BINAURAL models have a good fit for the changing trend of the sound pressure level curve. 
The proposed method has a better fitting performance in the sound pressure level curve of the left channel (see Fig. \ref{fig:sixcol}(b)).
The left channel sound pressure level curve of our model is above the MONO2BINAURAL. 
However, the fitting performance of the MONO2BINAURAL model is better than our method on the sound pressure level curve of the right channel (see Fig. \ref{fig:sixcol}(c)).
The curve of our model is roughly below MONO2BINAURAL. 
The reason for this phenomenon may be that the sounding object is located on the left side, so our model reduces the sound pressure level of the right channel. 
Similarly, our model increases the sound pressure level of the left channel, causing our curve to be above the MONO2BINAURAL. 
The sound pressure level difference curve can more intuitively observe the fitting effect of the model on spatial perception.
Compared with the MONO2BINAURAL model, our method has a better fitting performance in the sound pressure level difference curve, and most of them are above 0 dB scale (see Fig. \ref{fig:sixcol}(d)). 
It can be seen that the proposed method has a better spatial sound generation performance.

Fig. \ref{fig:sevencol} shows the experimental results of spatial perception on the YT-Music dataset.
The sound sources of the four instruments in the video are on the right side of the vision.   
The spatial perception direction of binaural stereo audio is biased to the right.
Fig. \ref{fig:sevencol}(b) and (c) show that SAGM performs better than MONO2BINAURAL in fitting the left and right sound pressure level curves. 
In Fig. \ref{fig:sevencol}(d), we utilize the sound pressure level difference curve to compare the magnitude and direction of spatial perception.
According to the magnitude of spatial perception, the order is the ground-truth, SAGM, MONO2BINAURAL from large to small. 
The binaural stereo audio generated by SAGM has the closest spatial perception magnitude to the ground-truth.
In the direction of spatial perception, the binaural stereo audio generated by MONO2BINAURAL does not show obvious directivity at the beginning (at the black ellipse). 
The spatial perception magnitude of binaural stereo audio generated by SAGM is smaller than the ground-truth, but it shows correct directionality (at the black ellipse).
At the orange ellipse, the spatial perception direction of MONO2BINAURAL is opposite to the ground-truth, while SAGM shows a weaker direction of spatial perception. 
At the yellow ellipse, both MONO2BINAURAL and SAGM show poor spatial perception. 
Generally speaking, the binaural stereo audio generated by SAGM is better than MONO2BINAURAL in the magnitude and direction of spatial perception. 
Therefore, SAGM has better generation performance and can generate more realistic binaural stereo audio.

\section{User Studies Discussion}
We organize two user studies on the FAIR-Play dataset to artificially evaluate the quality of the produced binaural stereo audio.
For the generalizability of the experiments, all subjects were not trained in stereo perception.
They voted on their intuitive feelings, under the premise of knowing the experimental background and rules.
Before the experiment began, a binaural stereo audio was provided to the subjects to test whether the headphones met the requirements.
\begin{figure}[t]
  \centering
   \includegraphics[width=0.8\linewidth]{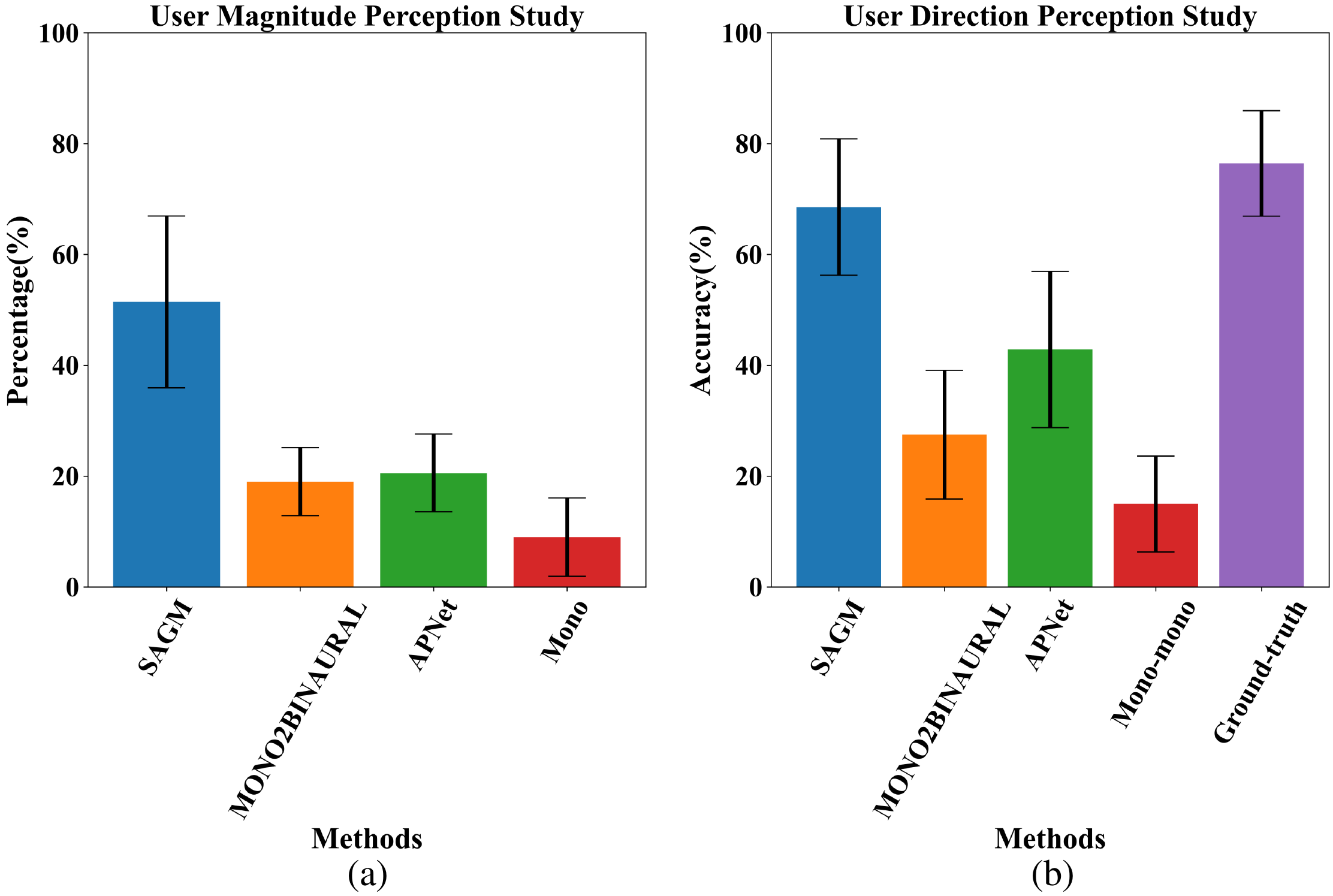}

   \caption{User studies on the FAIR-Play dataset.
   (a) is the user study of magnitude perception, while (b) is the user study of direction perception.}
   \label{fig:eightcol}
\end{figure}

The first user study is a spatial magnitude perception study.
The study samples are obtained by combining silent video with binaural stereo audio generated by different methods.
Firstly, the subjects are asked to watch a 10-second reference video with ground-truth audio.
Then, the subjects are asked to find the sample from the shuffled study samples that is closest in spatial sensation to the reference video.
We recruited a total of 14 subjects with normal hearing.
Each subject was required to vote on 50 groups of study samples.
Fig. \ref{fig:eightcol}(a) demonstrates the percentage of votes obtained by different methods.
We can see that the proposed SAGM received the highest percentage of votes.
Therefore, the binaural stereo audio generated by the proposed method provides a feeling closer to reality.

The second user study is a direction perception study.
We put ground-truth and audio generated by different methods in a group and shuffled them.
The subjects are asked to listen to them one by one and give the direction of the specified sound.
We determine the ground-truth of sound direction based on the position of a specified sound in the vision (left, middle, or right).
Specified sounds include various instruments and vocals.
Each subject needs to complete the direction selection of 20 groups of study samples.
Fig. \ref{fig:eightcol}(b) shows the accuracy of subjects choosing sound directions when listening to binaural stereo audio generated by different methods.
The ground-truth has the best directional indication, followed by the binaural stereo audio generated by our method.
Compared with other methods, the binaural stereo audio generated by the proposed method has a more accurate indication of sound direction.

User studies by us and others \cite{DBLP:conf/cvpr/GaoG19} have shown that user studies are highly subjective, which leads to a large degree of dispersion in experimental results.
The dispersion of results may be mitigated by training on subjects, but this will increase experimental cost and reduce generalization.
The proposed evaluation metric achieves a measure of the magnitude and direction of spatial perception in the temporal domain, which is a feasible alternative to user studies.

\section{Conclusion}
In this paper, we first propose a stereo generative adversarial network model named SAGM to generate binaural stereo audio from mono audio.
In SAGM, a shared spatio-temporal visual information is utilized to guide the generator and discriminator to generate and discriminate binaural stereo audio respectively.  
The proposed SAGM can generate more space-realistic audio through generative adversarial learning under guidance sharing. 
Experimental results show that the proposed SAGM outperforms other state-of-the-art methods in both quantitative and qualitative evaluations.
In addition, we propose a novel evaluation metric to demonstrate changes in spatial perception of audio.
The evaluation of spatial perception is a significant issue and has not been studied before. 
Hence, we believe that the proposed metric will serve as a new evaluation method for work in binaural stereo audio generation.
Finally, the proposed metrics can replace user studies to a certain extent, reducing the subjectivity and cost of manual evaluation.
Our method is based on spectrogram generation, and the knowledge provided to the generator is the coarse guidance information under the spectrogram features.
In the future, we will utilize visual guidance to generate binaural stereo audio directly on the waveform, thus avoiding tedious waveform-spectrogram transformations and mask calculations and providing data-level granular visual guidance.

\bibliographystyle{elsarticle-num}
\bibliography{tip}

\end{document}